\documentclass{WileyMSP-template}
\usepackage{xcolor}
\usepackage{ulem}
\usepackage{float}
\usepackage[utf8]{inputenc}
\usepackage[version=4]{mhchem}
\usepackage[separate-uncertainty=true]{siunitx}
\DeclareSIUnit\Molar{\textsc{m}}
\DeclareSIUnit{\wtpercent}{wt\%}
\DeclareSIUnit{\torr}{Torr}

\begin{document}
\sloppy

\hyphenation{a-na-ly-sis}

\pagestyle{fancy}
\rhead{\includegraphics[width=2.5cm]{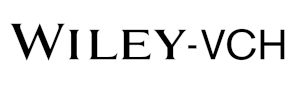}}

\title{Two-junction model in different percolation regimes of silver nanowires networks}

\maketitle

\author{J. I. Diaz Schneider}
\author{C. P. Quinteros}
\author{P. E. Levy}
\author{E. D. Martínez*}

\begin{affiliations}
J. I. Diaz Schneider, Pablo E. Levy E. D. Martínez*\\
Consejo Nacional de Investigaciones Científicas y Técnicas (CONICET), Argentina.\\
Instituto de Nanociencia y Nanotecnología (CNEA - CONICET), Nodo Bariloche.\\
Gerencia Física, Centro Atómico Bariloche, Comisión Nacional de Energía Atómica (CNEA), Av. Bustillo 9500, (8400) S. C. de Bariloche, Río Negro, Argentina. \\
*E-mail: eduardo.martinez@cab.cnea.gov.ar\\

C. P. Quinteros\\
Instituto de Ciencias Físicas (UNSAM-CONICET), Martín de Irigoyen 3100, San Martín (1650), Argentina.\\
\end{affiliations}


\begin{abstract}

Random networks offer fertile ground for achieving complexity and criticality, both crucial for an unconventional computing paradigm inspired by biological brains’ features. In this work, we focus on characterizing and modeling different electrical transport regimes of self-assemblies of silver nanowires (AgNWs). As percolation plays an essential role in such a scenario, we explore a broad range of areal density coverage. Close-to-percolation realizations (usually used to demonstrate neuromorphic computing capabilities) have large pristine resistance and require an electrical activation. Up to now, highly conductive over-percolated systems (commonly used in electrode fabrication technology) have not been thoroughly considered for hardware-based neuromorphic applications, though biological systems exhibit such an extremely high degree of interconnections. Here, we show that high current densities in over-percolated low-resistance AgNW networks induce a fuse-type process, allowing a switching operation. Such electro-fusing discriminates between weak and robust NW-to-NW links and enhances the role of filamentary junctions. Their reversible resistive switching enable different conductive paths exhibiting linear I-V features. We experimentally study both percolation regimes and propose a model comprising two types of junctions that can describe, through numerical simulations, the overall behavior and observed phenomenology. These findings unveil a potential interplay of functionalities of neuromorphic systems and transparent electrodes.

\end{abstract}

\section{Introduction}
\noindent Artificial neural networks (ANNs) enabled impressive development in many areas of science and technology, nowadays termed “artificial intelligence” \cite{lecun_deep_2015}. An ANN is a mathematical abstraction with a strong biological inspiration that emulates synaptic processes, implemented in codes (software) running on mass-produced electronic devices \cite{sebastian_memory_2020,markovic_physics_2020,Christensen2022}. As an alternative strategy, ANNs can be encoded in hardware, which would result both in the optimization of power consumption and the overall efficiency (miniaturization and velocity) as new calculation strategies and information processing methods are enabled \cite{Christensen2022,kuncic_neuromorphic_2021}.

For a hardware-based implementation of ANNs, one approach is to use ordered arrays of artificial synapses \cite{prezioso_training_2015}: their advantage is the possibility to assess the state of individual components. While the number of components that can be achieved using this top-down strategy may be reasonably large \cite{merolla_million_2014}, this overwhelming technological step is still insufficient if a realistic emulation of the connectivity of neurons is aimed. Moreover, it is desirable to incorporate a distinctive feature of biological neural tissues: intricate networks formed by enormously large amounts of simple components having an extremely high degree of interconnection \cite{chialvo_emergent_2010}. Thus, from qualitative and quantitative points of view, an alternative to ordered microcircuits is envisaged for the following breakthrough. 

Taking advantage of bottom-up strategies, a possible realization of ANNs that overcomes such limitations is the use of self-assembled networks of nanowires (NWNs) \cite{kuncic_neuromorphic_2021,diaz-alvarez_emergent_2019,milano_connectome_2022}.   
NWNs have been studied for two types of applications. The first one relies on their ability to act as transparent electrodes. Since the network is composed of highly interconnected metallic nanowires which still do not cover the full area, they have been intensively explored as both highly conductive and transparent materials \cite{khanarian_optical_2013,ye_metal_2014,zhu_convertible_2019}. 
Alternatively, the use of NWNs as a playground for neuromorphic processes is being extensively studied \cite{kuncic_neuromorphic_2021,diaz-alvarez_emergent_2019,milano_connectome_2022,zhu_convertible_2019,bellew_resistance_2015,milano_brain-inspired_2020,Zhu2023}. 

So far, low-density \textbf{percolated} (P) NWNs have been studied for neuromorphic purposes \cite{kuncic_neuromorphic_2021,diaz-alvarez_emergent_2019,milano_connectome_2022,bellew_resistance_2015,milano_brain-inspired_2020,Zhu2023}. Such NWNs display a relatively high pristine resistance that can be switched to lower resistance states by applying voltage pulses (referred as \textit{activation}). This processing, which has recently been applied successfully in signal recognition, handwritten digit classification, and chaotic time series prediction \cite{Zhu2023,Daniels2022,Milano2022}, owes its learning capability to the dynamic nature of the conductivity map across the network, which allows the projection of spatio-temporal mapped information onto a feature basis. 

Upon increasing the number of NWs per area, a high-density limit or \textbf{over-percolated} (OP) regime is reached. Samples obtained in this condition are comparable to those reported in terms of transparent electrodes \cite{Bellet2017,Gerlein2021,Manning2019}. Moreover, as we have shown in a previous report \cite{diaz_schneider_resistive_2022}, these OP samples also pose interesting electrical features suitable for neuromorphic applications.

In this report, by carefully covering the percolation curve, we demonstrate the multiplicity of electrical behaviors obtained, proposing a unifying model able to describe the overall spectrum of explored samples. Beyond the possibility of incorporating knowledge previously developed in research areas with diverging technological scopes, the existence of a unified model for the wide spectrum of NWN densities paves the way to plausible multifunctional devices.  

\section{Results and Discussion}

Metallic silver nanowires (AgNWs) coated with PVP were obtained following the standard polyol synthesis method \cite{Jiu2014}. Obtained NWs have characteristic length L = \SI{70 \pm 20}{\micro\meter} and diameter d = \SI{170\pm 30}{\nano\meter} (Synthesis E, see Table S1, Supporting Information), as determined from statistical analysis of multiple Scanning Electron Microscopy (SEM) and dark field optical microscopy images \cite{diaz_schneider_resistive_2022}. NWNs were fabricated on glass substrates following spin coating \cite{khanarian_optical_2013} and multi-deposition steps \cite{zhu_convertible_2019}.

\subsection{Percolation curve}
 
\begin{figure}[ht!]
    \centering
    \includegraphics[width = 1\textwidth]{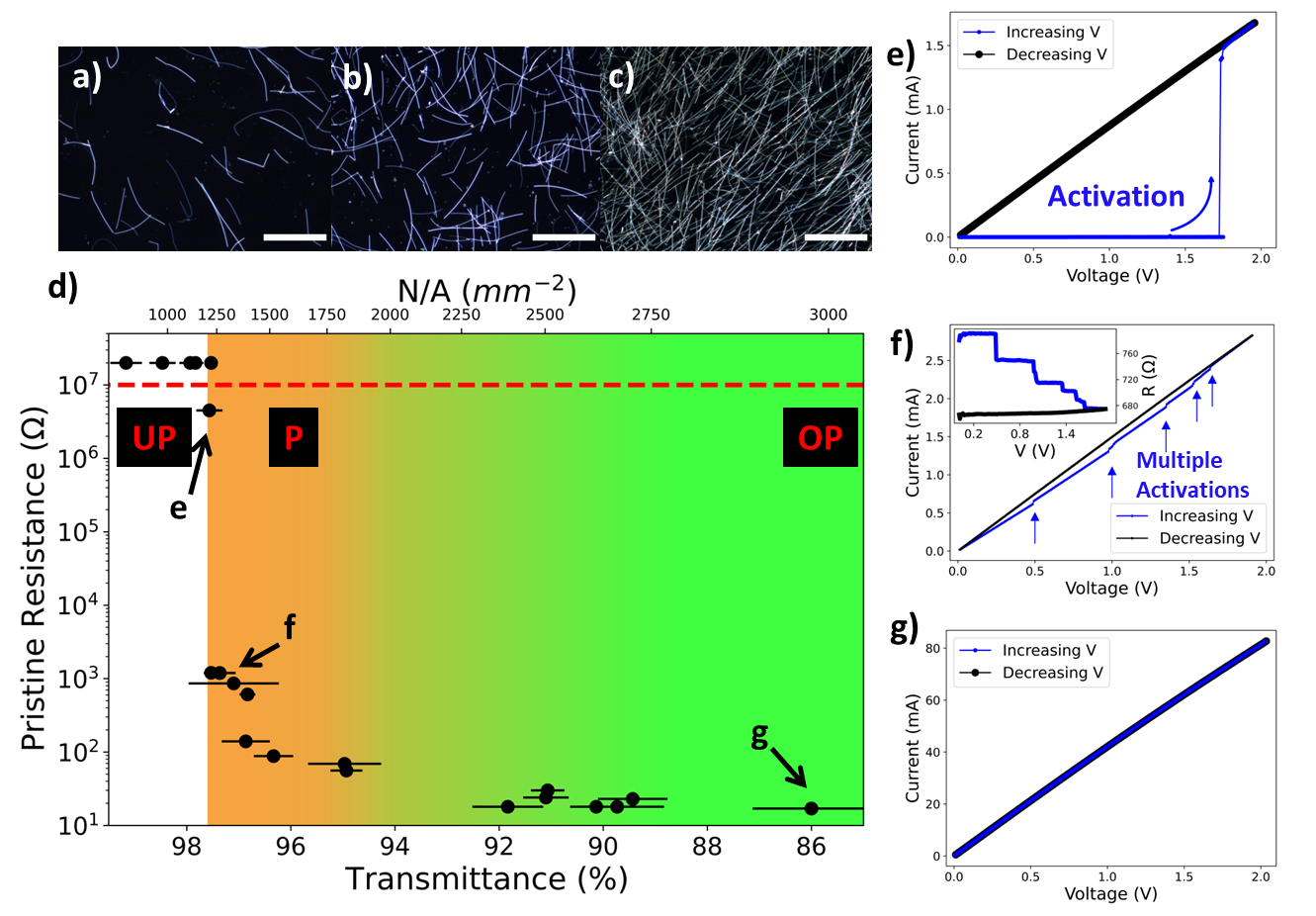}
    \caption{\textbf{Percolation curve of AgNWNs}. (a-c) Optical images of NWN films with three different AgNWs densities. Scale bar: \SI{100}{\micro\meter} (d) 4-wire pristine resistance as a function of transmittance (bottom axis) corresponding to different AgNWs densities (top axis). Dots over the red dashed line correspond to resistance values greater than the measuring capability. (e-g) Current as a function of voltage obtained during the first run for three samples of different NW densities indicated in (d), in the P (e,f) and OP (g) regimes, respectively.}
    \label{fig:percolation}
\end{figure}

A percolation curve can be obtained by varying the number of deposition steps (see Figures \ref{fig:percolation}a-d). More steps determine a higher substrate coverage or areal density as depicted in the dark field optical microscopy images of Figures \ref{fig:percolation}a-c. To quantify the coverage we use the optical transmittance at \SI{550}{\nano\meter} (higher sensitivity of the human eye) as commonly used for transparent electrodes \cite{Jeong2018,Goliya2019,Li2019}.
In turn, the areal density ($\mathrm{\frac{N}{A}}$) is obtained from the transmittance using a calibration previously conducted (shown in S1, Supporting Information) \cite{nirmalraj_manipulating_2012}.  
Figure \ref{fig:percolation}d shows the obtained percolation curve. In agreement with previous reports \cite{chung_solution-processed_2012}, a higher substrate coverage implies a lower optical transmittance and a concomitantly lower pristine resistance. It is thus possible to define a critical coverage fraction (represented in Figure \ref{fig:percolation}d as a vertical line at a transmittance value of 97.6\%) as the boundary between an \textbf{under-percolated} (UP) regime, in which the electrical resistance exceeds the measuring limit of the instrument (indicated as a red dashed line), and the \textbf{percolated} (P) regime in which the resistance is measurable (below 10 M$\Omega$). 

Close to that boundary, with a high-resistance state as the pristine condition, the P regime is known to require activation (Figure \ref{fig:percolation}e) to switch to low-resistance states \cite{kuncic_neuromorphic_2021,diaz-alvarez_emergent_2019,milano_connectome_2022,bellew_resistance_2015,milano_brain-inspired_2020,Zhu2023}. When $\mathrm{\frac{N}{A}}$ increases, the electrical behavior progressively evolves into completely different electrical responses (see Figures \ref{fig:percolation}f-g). Illustrating this evolution, Figure \ref{fig:percolation}f presents multiple activation steps, a phenomenology poorly explored in similar systems. Moreover, upon further increasing $\mathrm{\frac{N}{A}}$, an OP regime displaying a pristine I-V linear response is eventually reached (Figure \ref{fig:percolation}g). 

In the following, we show the characteristic electrical response of several samples in the P and OP regimes. As the topology of the AgNWNs plays a fundamental role and depends on the random arrangement of AgNWs during their self-assembly, the samples were studied to identify representative features that were common to all samples in each regime. 

\subsection{Percolated regime}

\begin{figure}[ht!]
   \centering
   \includegraphics[width = 0.7\textwidth]{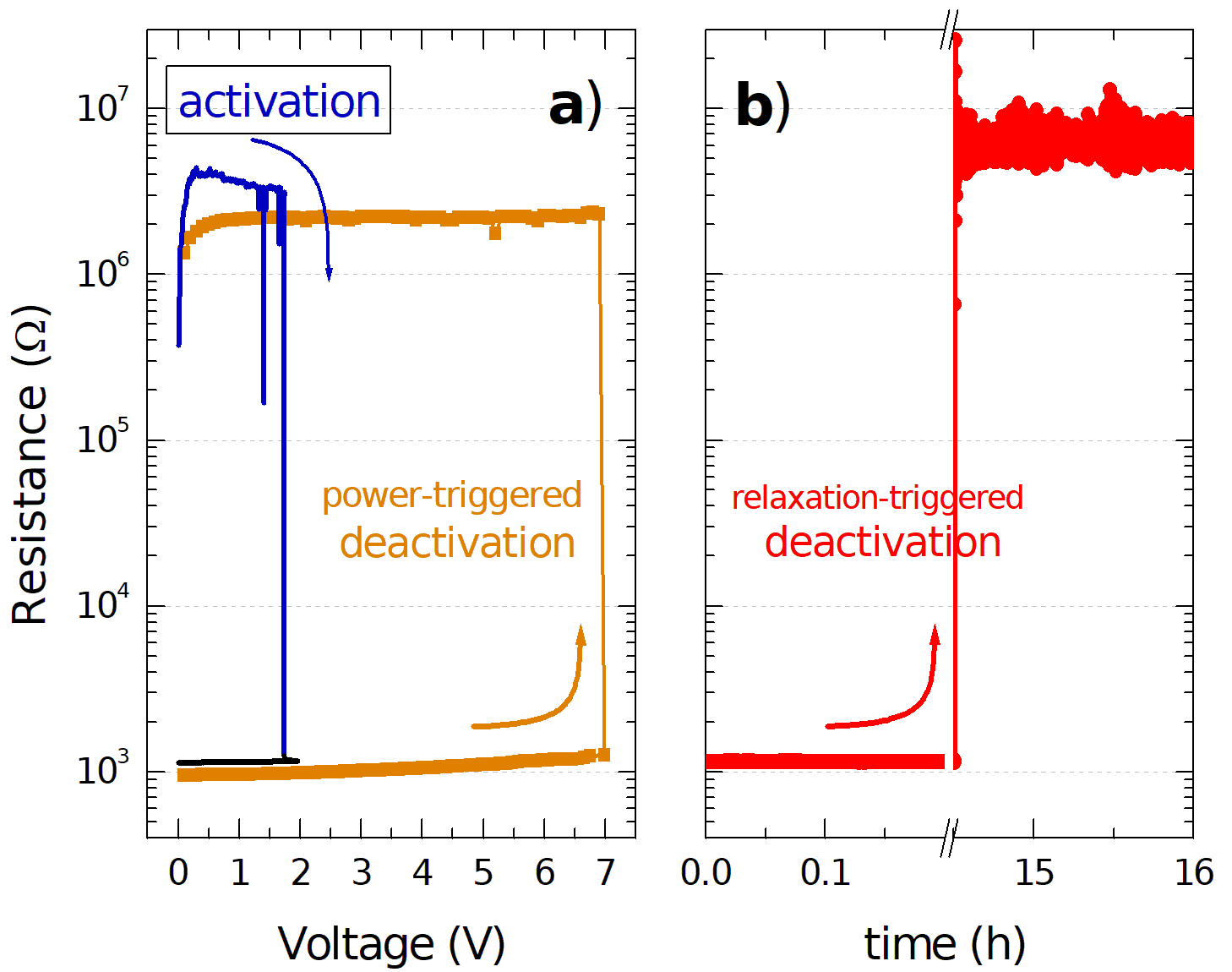}
   \caption{\textbf{Deactivation mechanisms observed in a P sample.} These measurements correspond to the same P sample presented in Figure \ref{fig:percolation}e. (a) Resistance as a function of voltage. Two voltage sweeps are included. In the first run, the pristine state becomes more conductive, usually referred to as activated (same data as in Figure \ref{fig:percolation}e). During the second run, power-triggered deactivation is observed. (b) Resistance as a function of time recorded at \SI{1}{\milli\volt}, demonstrating relaxation-triggered deactivation is also possible.}
   \label{fig:F2}
\end{figure}

Samples prepared in the P regime display an electrical transport behavior that was explored in great detail by other groups in an attempt to exploit its capacity for neuromorphic computation \cite{kuncic_neuromorphic_2021,diaz-alvarez_emergent_2019,milano_connectome_2022,bellew_resistance_2015,milano_brain-inspired_2020,Zhu2023}. In an initial high-resistive pristine state, percolated samples can switch between different states by activating or deactivating junctions within the assembly. Figure \ref{fig:F2} shows that deactivation occurs either by sweeping the externally applied electrical stimulus (which, in turn, increases the dissipated power) or by allowing the system to relax. This duality is at the core of many proposals by multiple authors to process temporal information \cite{Zhu2023,Daniels2022,Milano2022}. 

\begin{figure}[ht!]
   \centering
   \includegraphics[width = 0.5\textwidth]{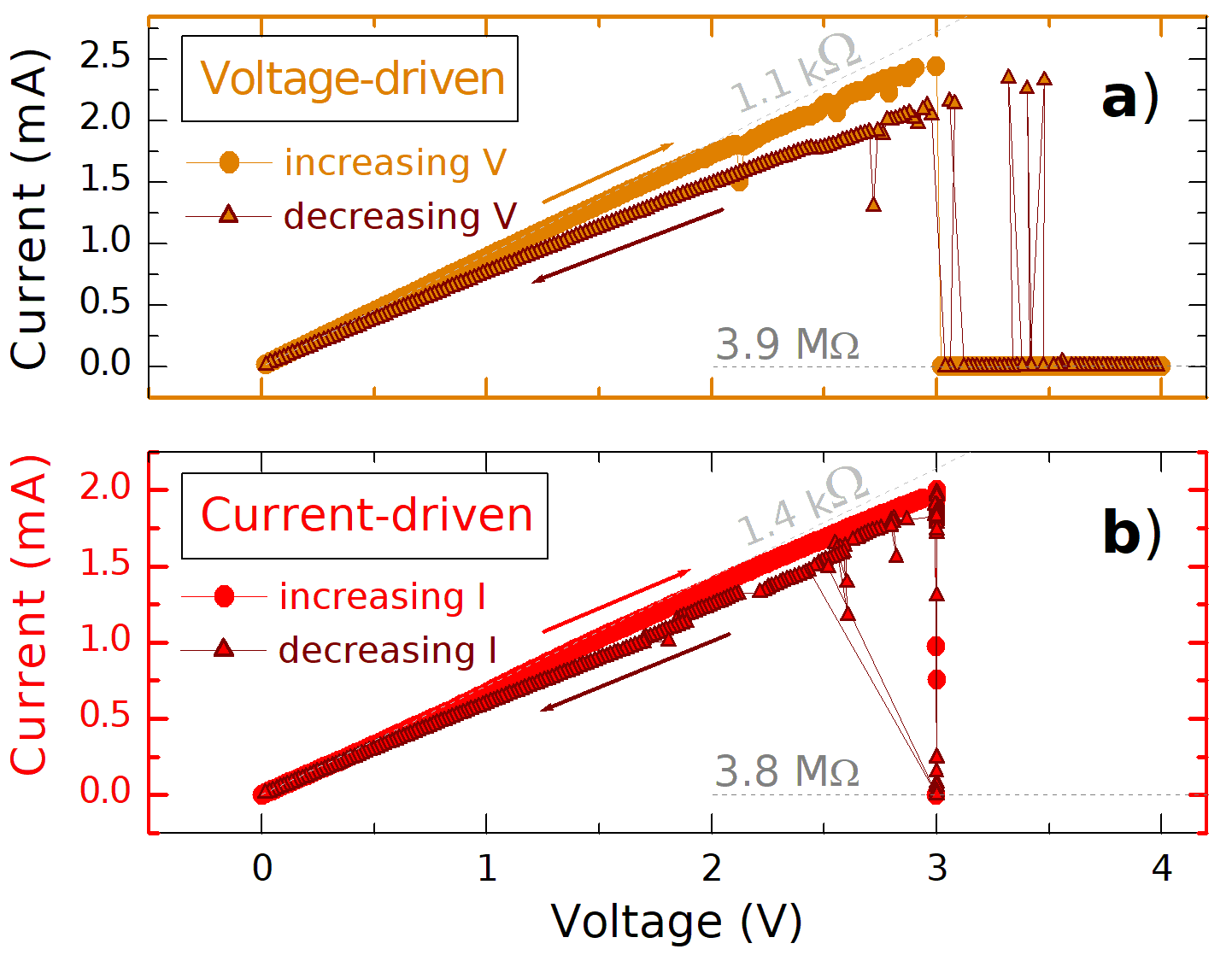}
   \caption{\textbf{Reversible I-V behavior observed in a P sample.} (a) Voltage-driven and (b) current-driven I-V curves were measured on the same percolative sample (also the same as in Figure \ref{fig:percolation}e and Figure \ref{fig:F2}). The electrical response depicted in (a) and (b) qualitatively coincides regardless of the driving stimulus and whether it is increased or decreased. Both cases are plotted using the same combination of I-V axes to emphasize their similarities.}
   \label{fig:zig-zag-P}
\end{figure}

Interestingly, upon electrically cycling the sample of Figure \ref{fig:F2}, a completely different behavior is obtained. Figure \ref{fig:zig-zag-P} comprises I-V curves displaying a low-resistance (LR) state at low voltages with a switch to a high-resistance (HR) state at higher voltages. This phenomenology occurs regardless of the driving stimulus, V or I, displayed in Figures \ref{fig:zig-zag-P}a and \ref{fig:zig-zag-P}b, respectively. While increasing the driving stimulus the switching occurs from LR to HR, decreasing it produces a mirror response: changing from HR to LR state at approximately the same I-V condition.   

Although different from the salient features reported for AgNWNs studied for neuromorphic implementations, this reversible I-V behavior, shown in Figure \ref{fig:zig-zag-P} for positive polarity, will prove to be ubiquitous in other zones of the percolative curve. 

So far, close-to-percolation samples, similar to those studied by other authors \cite{kuncic_neuromorphic_2021,diaz-alvarez_emergent_2019,milano_connectome_2022,bellew_resistance_2015,milano_brain-inspired_2020,Zhu2023}, have been discussed. Heading toward the OP regime of interest, here we would like to emphasize the intermediate responses obtained upon increasing $\mathrm{\frac{N}{A}}$. As indicated in Figure \ref{fig:percolation}f, samples with a higher degree of coverage exhibit multiple successive activation steps. This observation suggests that steady states are achievable regardless of the monotonous increase in voltage (see inset in Figure \ref{fig:percolation}f). Either specific values of V are required for the progressive resistance reduction (Figure \ref{fig:percolation}f), or the changes occur after a certain time. Additional measurements also indicate that the persistence of the achieved resistance state strongly depends on the repetition (Figure S2a, Supporting Information), amplitude (Figure S2b, Supporting Information), and duration (Figure S3, Supporting Information) of the voltage pulses. This points out an interplay between the external driving force and the timescale, the latter probably through the dissipated power, in turn, dependent on the network connectivity which sets the current flow in different NWs.

\subsection{Over-percolated regime}

In a previous contribution \cite{diaz_schneider_resistive_2022}, we presented and characterized AgNWNs focusing on the role of environmental conditions. Those samples belong to the regime labeled as OP in this work. In that former communication, we identified several key aspects of the electrical response, such as a pristine low resistance state, and the need for an irreversible operation to enable further switching. Here, we deepen our understanding of the underlying phenomenology by complementing the macroscopic electrical behavior with microscopic techniques and systematically analyzing a broader variety of samples. 

\begin{figure}[ht!]
    \centering
    \includegraphics[width = \textwidth]{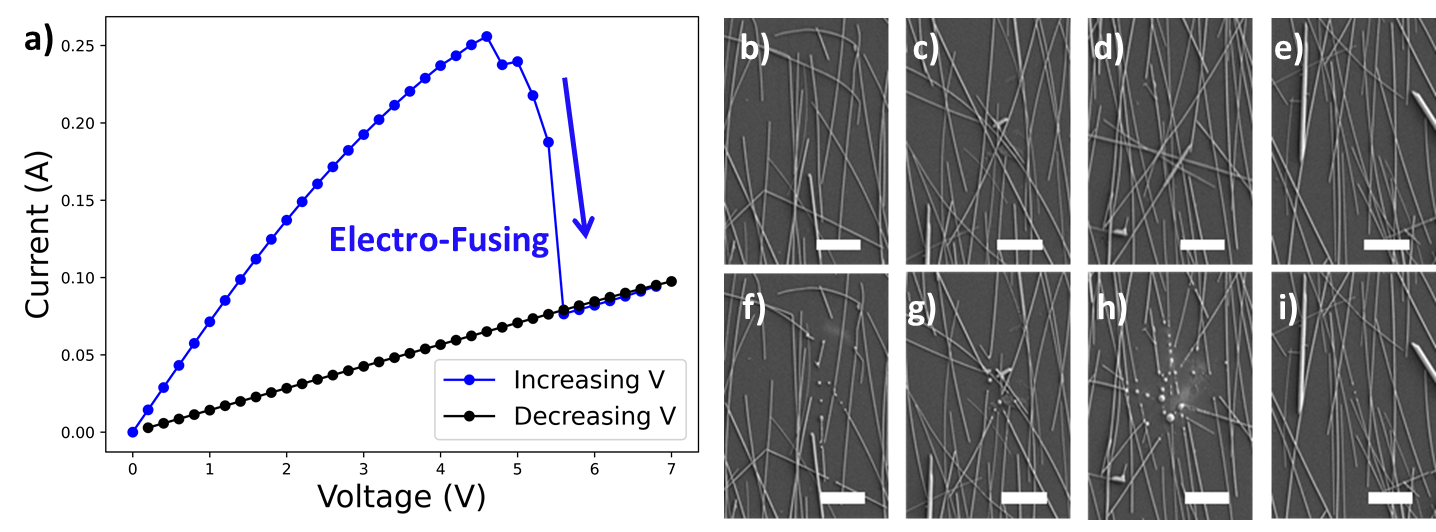}
    \caption{\textbf{Typical electro-fusing (EF) process of OP samples}. (a) Current as a function of voltage (I-V) dependence during the EF operation including SEM images of the (b-e) pristine and (f-i) post-EF states. Scale bar corresponds to \SI{5}{\micro\meter}.}
    \label{fig:F3}
\end{figure}

As shown in Figure \ref{fig:percolation}g, pristine OP samples show low-resistance ($\sim$ \num{d1} \unit{\Omega}) describing a linear I-V dependency up to moderate values of voltage (below \SI{2}{\volt}). Upon increasing the applied voltage, the response departs from linearity typically for currents exceeding $\sim$ \SI{200}{\milli\ampere} (see Figure \ref{fig:F3}a). Considering the linearity of the low-stimulus I-V as an indication of a metallic-like nature of the percolative path, the smooth R increase above \SI{200}{\milli\ampere} can be related to a thermal effect of the dissipated power. A subsequent marked current decline is observed. An induced fuse-type process, hereafter termed electro-fusing (EF), produces an irreversible switching to a high-resistance state referred to as R$_{\mathrm{EF}}$ (Figure \ref{fig:F3}a). 

The EF appears gradual (multiple steps) or abrupt (single step) depending on the dynamics of the system (which, in turn, could 
evidence an avalanche-like process \cite{hochstetter_avalanches_2021,dunham_nanoscale_2021})
and the sweeping protocol \cite{kholid_multiple_2015}. 
The EF process comprises a distinctive feature of the OP samples. Figure \ref{fig:F3}a depicts a typical current-voltage (I-V) dependence during the previously described stages. Figures \ref{fig:F3}b-e and \ref{fig:F3}g-j show SEM images obtained on different regions before and after the realization of the EF, respectively. Other examples of EF can be found in our previous work \cite{diaz_schneider_resistive_2022}.

As different from activation (the protocol usually used to form a conductive path along an otherwise insulating media, thus exhibiting a concomitant decrease in R), here EF refers to an increase of R related to the breakdown of conductive paths, resembling a fuse device, as indicated by the SEM images in Figure  \ref{fig:F3}b-j. AgNWNs become disrupted due to excessive dissipated heat produced by the high current flow \cite{milano_electrochemical_2024}. We observe spheroidization of the AgNWs that is typically reported at temperatures above \SI{300}{\degreeCelsius} due to Plateau-Rayleigh instability \cite{Khaligh2013,Lagrange2017,Deignan2017}, in agreement with the formation of nanogaps as reported by Milano \textit{et al.} \cite{milano_brain-inspired_2020,milano_electrochemical_2024}.

EF can be interpreted as an electrical strategy to reduce the connectivity of the AgNWN allowing tuning the sample's density of conductive paths. This is comparable to a controlled shift in the percolation curve (Figure \ref{fig:percolation}) from highly dense networks to those in the P regime.

After this initial and irreversible EF, ulterior V cycling depicts a reversible linear I-V response, including both polarities. Samples with R$_{\mathrm{EF}} \sim$ \num{d2}-\num{d5} \unit{\ohm} are obtained after this fusing process \cite{diaz_schneider_resistive_2022}. The I-V response is further explored for voltages below \SI{2}{\volt}. 

\begin{figure}
    \centering
    \includegraphics[width=\textwidth]{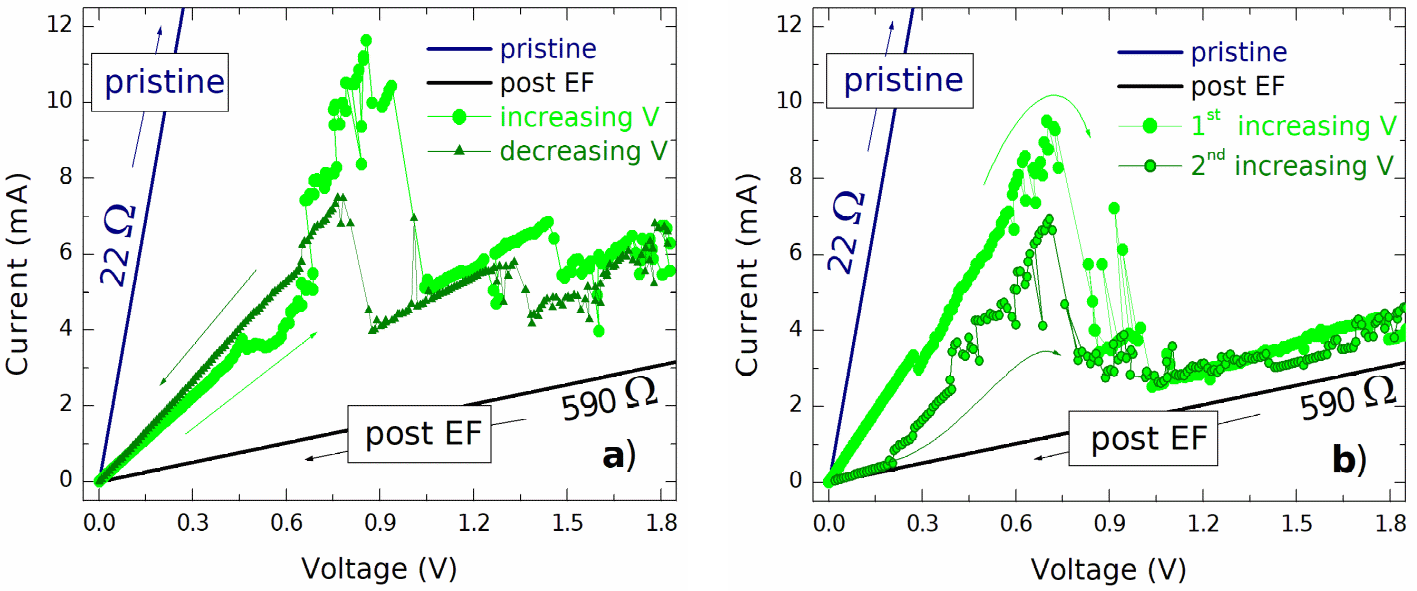}
    \caption{\textbf{Reversible I-V behavior in an OP sample post EF using two protocols:} (a) a first increasing V-ramp followed by a decreasing one, and (b) two successive increasing V-sweeps.}
    \label{fig:IV_sweeps}
\end{figure}

During several cycles, the response remains constant at R$_{\mathrm{EF}}$ (indicated in Figure \ref{fig:IV_sweeps} as 'post EF' state). This stage is considered to be key in the subsequent electrical response. Eventually, the samples become responsive depicting a switching operation (Figure \ref{fig:IV_sweeps}). To test the nature of this observed switching, Figure \ref{fig:IV_sweeps} includes the electrical response upon two different protocols. On the one hand, Figure \ref{fig:IV_sweeps}a shows a low-resistance state which eventually becomes higher upon increasing the stimulus $\sim$ (0.9 V; 10 mA). Reduction of the externally applied stimulus (V, in this case) consistently displays a high resistance state for the higher V range resuming a low-resistance state at low V values. Complementary, Figure \ref{fig:IV_sweeps}b comprises the electrical response of the same OP sample upon two successive forward runs. While the first run is similar to the phenomenology described in Figure \ref{fig:IV_sweeps}a, the second run behaves differently: it departs from the origin (0 V; 0 A) in a high resistance state compatible with the remaining state of the previous run. The comparison between the two protocols allows us to spot the persistence of the achieved high resistance state which could be lost depending on the specific routine applied to the system.  

The peculiar switching that we have formerly introduced in Diaz Schneider \textit{et al. }\cite{diaz_schneider_resistive_2022}, and we are revisiting in this communication, demonstrates to be ubiquitous in different regimes of the percolation curve (see Figure \ref{fig:zig-zag-P} and \ref{fig:IV_sweeps}) but also in several ($\sim$50) OP samples. The statistical analysis performed on multiple samples regardless of the electrodes' position, NWs geometry, and measurement strategy is shown in Supporting Information (Figures S9 and S10) to emphasize the representativeness of this behavior.
The common features after EF is the presence of two or more linear regimes in the I-V curves (see the double logarithmic plot in Figure S5), with switching operations occurring at a condition labeled as V$_{\mathrm{S}}$ (see Figure S4) between an LR state, at V $<$ V$_{\mathrm{S}}$, and an HR state, at V $>$ V$_{\mathrm{S}}$. Both V$_{\mathrm{S}}$ and the resistance change, $\Delta \mathrm{R} = \mathrm{HR}-\mathrm{LR}$, are well-defined as indicated by the histograms in Figure S6.  
 
As shown in the percolation curve (Figure \ref{fig:percolation}d), the so-called OP regime represents an extreme case of percolative samples, in which $\mathrm{\frac{N}{A}}$ is so dramatically increased that metallic-like behavior governs the pristine state until some of the main connections are disrupted by the EF operation. Next, a switching operation triggered by an electrical stimulus and determined by the elapsed history of applied stimuli and dissipated power is obtained. 

As indicated in the case of P samples, time also plays a role, most likely through the dependence of the heat produced due to the electrical power dissipated during the associated timescale. We refer to the Supplementary Information (Figures S7 and S8) for additional measurements performed on OP samples in the time domain.

\section{Modelling and simulation} \label{model}

We have expressed so far the characteristic features of the electrical transport in AgNWNs depending on the areal density ($\mathrm{\frac{N}{A}}$) of the assembly. Once the percolation has been achieved, two main regimes have been highlighted. In the well-documented percolated (P) regime (close to the percolation threshold), the assembly of AgNWs depicts a pristine highly resistive state. Applying electrical stimuli, those samples switch to an LR state. Resistance can be switched back to an HR state by either electrically cycling the sample or letting it vanish during a convenient timescale. In the P regime, the mechanism behind the reversible switching is commonly explained through the electrochemical formation of a conductive filament within the PVP-mediated junctions between AgNWs, producing a dramatic reduction in their contact resistance \cite{kuncic_neuromorphic_2021,Wang2019,Yang2020}. The formed nanofilament is thermodynamically unstable and labile.  Immediately after the external stimuli are interrupted a relaxation process starts to act, leading to the filament breakdown and the recovery of the HR state. In addition, excessive power also triggers the disruption of the nanofilament and the switching to the HR state. In this simple picture, the formation and rupture of nanofilaments within junctions lie behind the so-called \textit{activation} and \textit{deactivation} processes characteristic of P samples \cite{Wang2019}. The dynamics of these processes give AgNWNs in the P regime a fading memory capability, enabling temporal processing of multiple spatial inputs \cite{Milano2022}. 

Upon further increasing $\mathrm{\frac{N}{A}}$, the properties of the AgNWs assemblies become so markedly different that we refer to them as over-percolated (OP) samples. Their pristine state is highly conductive and exhibits a linear I-V response. Noteworthy, surpassing an amount of dissipated power, the pristine low-resistance increases markedly, and reversible switching is enabled. Instead of a resistance reduction, as it were the case in P samples, the irreversible step referred to as electro-fusing comprises a resistance increase of the pristine state. It is worth mentioning that after the corresponding EF process in the OP case, some of the switching characteristics are qualitatively similar to those found in the P ones (i.e. compare Figures \ref{fig:zig-zag-P} and \ref{fig:IV_sweeps}a).

Therefore, the resistive switching properties depend on the mere concentration of NWs within the network. Since all the AgNW assemblies are produced from the same precursors following a similar procedure, the key to unveiling the two phenomenologies should rely on the same constituents. 

Many studies attempted explanations for the resistance modulation of the NWNs, in general \cite{khanarian_optical_2013,ye_metal_2014,song_nanoscale_2014,rocha_ultimate_2015-2,hwang_influence_2016,sannicolo_electrical_2018, bellet_metallic_2019}, and the interwire junctions, in particular \cite{bellew_resistance_2015}. For samples that we would classify as belonging to the P regime, Nirmalraj \textit{et al.} \cite{nirmalraj_manipulating_2012} argued that the voltage threshold for activation corresponds to a power law of the areal density V$_{\mathrm{T}}=(\mathrm{\frac{N}{A})^n}$ where the exponent $n$ is a negative number that depends on the distance between electrodes. The higher $\mathrm{\frac{N}{A}}$, the lower V$_{\mathrm{T}}$ required to activate the first conductive path. They rationalized their experimental findings as the electrical breakdown of leaky-capacitive interwire junctions. For highly-dense AgNWs samples lying in the OP regime, as those used in transparent electrode applications, it has been reported that numerous post-processing operations affect the junctions, further decreasing the already low overall resistance of the assembly \cite{Garnett2012,Jin2018,Chung2020,Zeng2022}. 

Here, we propose an explanation applicable to the whole percolation curve (i.e. both P and OP regimes) based on the presence of two types of junctions. On the one hand, a metallic unalterable-type of junction
(J$_\mathrm{I}$) is comparable to the metallic resistance of the AgNWs themselves. The second one is a switchable filamentary junction (J$_\mathrm{II}$) which explains the switching ability of the system. The two types of junctions could have the same origin: wire-PVP-wire junctions that could either be robust and stable due to the presence of a metallic bridge or a welded union between the two NWs (J$_\mathrm{I}$) or insulating PVP junctions (J$_\mathrm{II}$) which, upon electrical stimuli, are suitable for the growth of conductive filaments that connect the two Ag cores and enable additional conductive paths for the current to flow within the NWNs.

Availability of robust unaltered junctions (J$_\mathrm{I}$)  
rely on the presence of remaining Ag species (Ag(I)) in the PVP layer and the dependence on the standard electrode potential of metal nanocrystals with their characteristic dimension (NWs' diameter). Regarding this last point, in Ag nanoparticles, the redox potential of Ag has been shown to shift negatively (to more cathodic values) from that of the bulk metal, being the magnitude of the potential shift inversely proportional to the particle size \cite{Plieth1982}. Additionally, the spontaneous welding of AgNWs junctions has been recently demonstrated experimentally by Lee \textit{et al.} \cite{Lee2018}. Therefore, we include in our model that after the self-assembly of the AgNWs, electrochemical processes at the nanoscale take place, leading to a significant population of welded and already-formed filament junctions. The presence of ionic species (e.g. Ag(I)) necessary for the filament formation within the switchable junctions (J$_\mathrm{II}$) also explains their switching properties.

\begin{figure}[hb!]
    \centering
    \includegraphics[width = 0.8 \textwidth]{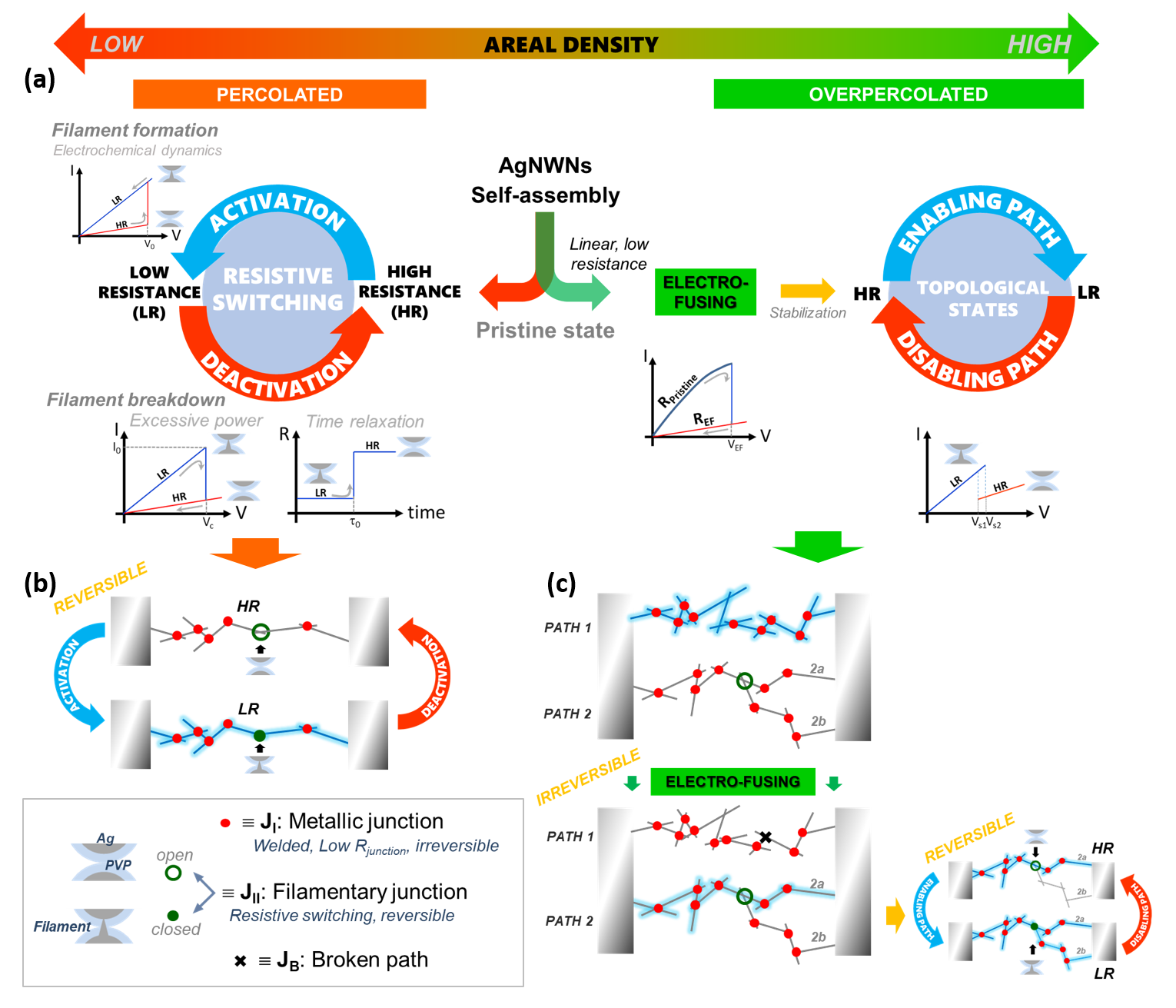}
    \caption{Scheme of the proposed model. (a) Self-assemblies of AgNWNs with areal density (N/A) are classified as percolated (P) or over-percolated (OP). In the P regime, the pristine state is highly resistive. Resistive switching (RS) to a low resistance state (LR) is triggered by the formation of a conductive filament within NWs junctions. Its breakdown may occur by excessive power dissipation or by spontaneous disintegration, leading to the return to a high resistance state (HR). In the OP regime, AgNWNs display low resistance. An irreversible electro-fusing process, followed by multiple voltage sweeps for stabilization allows RS by enabling or disabling percolating paths across the network. A diagram illustrating the conductive paths and their modifications is presented in (b) for NWNs in the P regime where the reversible activation and deactivation of a J$_\mathrm{II}$ junction produces the RS; (c) for NWNs in the OP regime. Here, a generic NWN may contain two characteristic percolative paths. In path 1, a purely metallic path with low resistance dominates the electrical conduction (blue glow) and is responsible for the initial LR state. Under the EF process, irreversible damage in some junctions or wires across path 1 produces a sudden increase in the resistance, enabling path 2 as the new lower resistance path. The following switching of the filamentary junction (indicated by the black arrow on top), produces the commutation between two topological resistance states: 2a and 2b. In path 2a, a percolative path is formed by purely metallic junctions (red dots) while the activation of an opened filamentary junction (green open circle) can enable a parallel track for conduction (path 2b). }
    \label{fig:scheme}
\end{figure}

Figure \ref{fig:scheme}a represents a sketch of the phenomenological characteristics of the AgNWNs depending on the areal density. For AgNWNs in the P regime, activation, and deactivation processes are behind the switching properties, labeling the formation and breakdown of filaments between NWs. For the OP regime, the pristine low resistance state is affected by the EF process, which abruptly modifies the network topology, leading to an HR state. The remaining network is able to switch between resistance states by enabling or disabling new percolation paths also governed by filamentary junctions that switch between open (no filament) or closed (formed filament) states. Schematic I-V curves of the different processes are shown to assist interpretation.

Considering the network topology, in the P regime the resistance of the percolating path depends on the state of a single (or a reduced number of) filamentary J$_\mathrm{II}$ junctions (Figure \ref{fig:scheme}b). In the OP regime, the electrical properties of samples are based on the two types of proposed junctions. In the pristine condition, the resistance is as low as the one corresponding to a unique Ag NW ($\sim$ \SI{85}{\ohm}). This could be rationalized as the presence of numerous parallel and serially connected conductive paths linking the electrodes, comprised of multiple NWs and robust metallic interwire junctions. Being the resistance of those metallic welded or already-formed filamentary junctions negligible compared to the resistance of the NWs themselves, the EF operation provokes the breakdown of some NWs and filaments formed at the junctions, as indicated in Figure \ref{fig:scheme}c and explicitly shown in Figure \ref{fig:F3}f-i. Disruption of the initially dominant metallic, low-resistance paths (e.g. path 1 in Figure \ref{fig:scheme}c), affects the distribution of current flowing through the remaining percolative paths (e.g. path 2 in Figure \ref{fig:scheme}c) that relay on the state of filamentary junctions.  
The switching of these critical J$_\mathrm{II}$ junctions enable/disable parallel percolating paths (e.g. path 2b) that result in the reduction/increase of the overall resistance. The ability to switch to a higher resistance state that could be reversed by a combination of externally applied electrical stimuli and timescale is successfully rationalized within this proposed framework. Moreover, this model is also reasonable for P samples by just modifying the density of NWs.

To test our hypothesis we performed simulations that considered all the key ingredients identified to play a role. The simulations comprise a set of wires randomly oriented within a frame in a scale proportional to the measured samples (70 $\mu$m-length NWs and 1 mm-apart electrodes). The two lateral extremes of the frame are considered to be the equipotential surfaces of the electrodes. At each NW crosspoint, there is a junction that could be either J$_\mathrm{I}$ or J$_\mathrm{II}$. The overall electrical response of the pristine NWN would depend on the ratio between the population of the two types, among other parameters, as we discuss in the following. 
    
\begin{figure}[ht!]
    \centering
    \includegraphics[width = \textwidth]{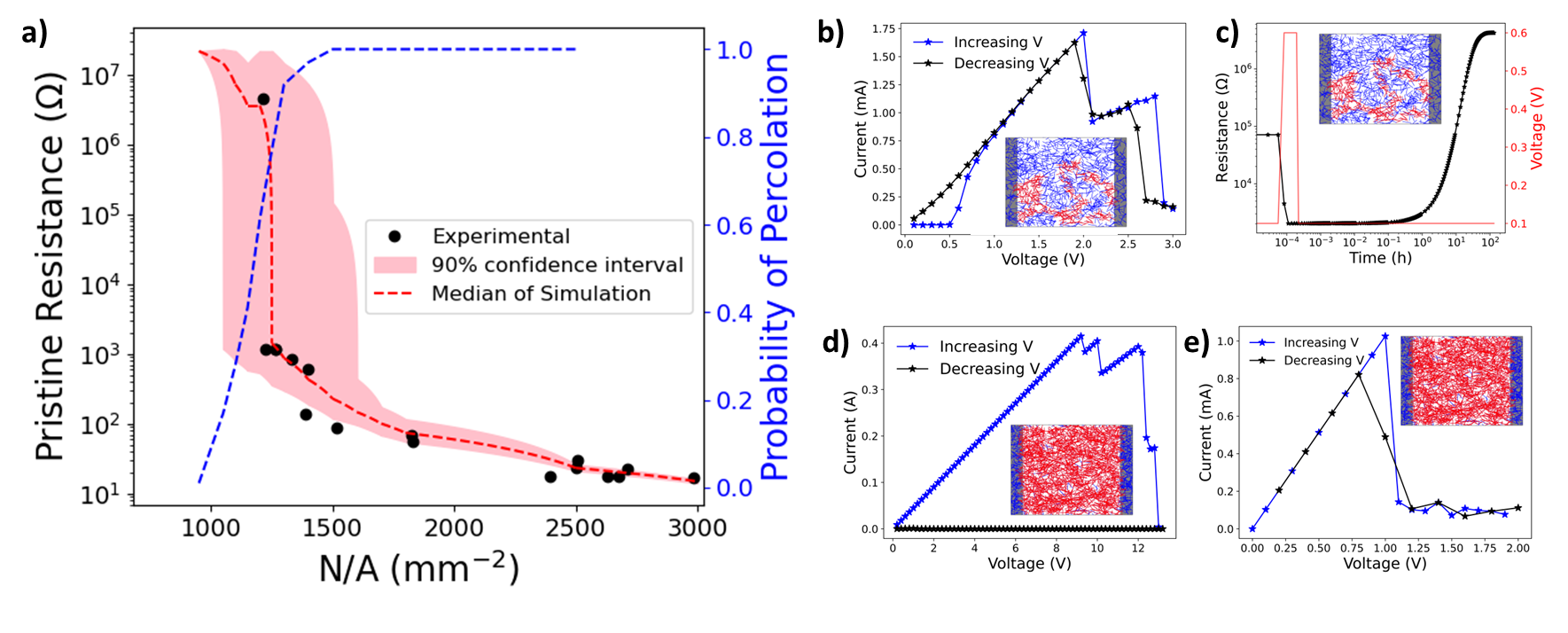}
    \caption{\textbf{Simulations performed using the two-junction model.} a) Percolation curve having 96$\%$ of metallic junctions and R$_\mathrm{fo}=$  \SI{8.5}{\mega\ohm}. 100 simulations were performed for each NW density. b) Simulated I-V curve and c) activation and relaxation processes for a 1200 AgNWs/mm$^2$ network. Simulated I-V curves for samples in the OP regime showing d) the initial state and the electro-fusing process, and e) I-V curves after EF for a 2500 AgNWs/mm$^2$ network. The insets are simulated NWNs where NWs indicated in red are the ones that form the percolated pathways between the electrodes.}
    \label{fig:sim}
\end{figure}

The resistance of the NWNs is represented by the sum of a fixed term (representing the electrodes' contact resistance and the contribution of the NWs, R$_\mathrm{w}$) and the terms corresponding to all the junctions. Nanowires can break if they exceed the critical power (P$_\mathrm{cw}$). J$_\mathrm{I}$ junctions exhibit linear response with a resistance of approximately \SI{0.1}{\milli\ohm} each (assuming a junction of 150 x 150 nm$^2$ and a thickness of \SI{10}{\nano\meter} using bulk silver resistivity). J$_\mathrm{II}$ junctions have an initial resistance R$_\mathrm{fo}$ related to the unformed filament. Once the voltage across the junction terminals exceeds a critical voltage difference (V$_\mathrm{c}$), a filament begins forming, gradually reducing the resistance until it reaches a final value of R$_\mathrm{fc}$ when fully formed. This filament may degrade over time due to thermodynamic instability or by exceeding the critical power threshold (P$_\mathrm{cf}$). Parameters such as the ratio between the population of the two types of junctions and the power threshold for the onset of the competition among them and others are custom-defined. The results are shown in Figure \ref{fig:sim}.

The quantity of AgNWs was varied, and the simulated network resistance was analyzed for different values of R$_\mathrm{fo}$ (Supporting Information). For R$_\mathrm{fo}$ $>>$ R$_\mathrm{w}$, there exists a critical fraction of nanowires for which the electrical resistance of the network undergoes an abrupt change (see Figure S11, Supporting Information). Below this concentration threshold, the network predominantly exhibits high resistance values, while above it, the resistance value drops sharply. Upon reduction of R$_\mathrm{fo}$ values, this change in NWN resistance diminishes until it disappears for R$_\mathrm{fo} \sim$ R$_\mathrm{w}$, in agreement with the expected smooth change of resistance for a network with an increasing number of series/parallel conducting paths. Additionally, as the percentage of filamentary junctions increases, the change in NWN resistance shifts towards higher wire densities. 

In Figure \ref{fig:sim}a, the percolation curve for R$_\mathrm{fo}=$ \SI{8.5}{\mega\ohm} and 96$\%$ metallic junctions is depicted. The dashed line represents the median value of 100 simulations, while the shaded region corresponds to simulations where resistances fall within the 5$\%$ and 95$\%$ percentiles. Experimental results, shown as black points, are plotted for comparison. The simulation can elucidate the concentration threshold between resistance values in the percolation curve. Noteworthy for lower-density values, achieving an electrically percolating network becomes more challenging, consequently reducing the likelihood of obtaining networks with higher resistance. Using a P$_\mathrm{cw}$ of \SI{30}{\milli\watt} (a value consistent with Milano's findings \cite{milano_brain-inspired_2020}), P$_\mathrm{cf}$ of \SI{1}{\milli\watt}, V$_\mathrm{c}=$ \SI{0.4}{\volt}, and a ratio between open and closed J$_\mathrm{II}$ resistance equal to 10000, the electrical behavior of networks with various nanowire densities were simulated. Details regarding the simulation of electrical behavior are provided in the Supporting Information. 

In Figure \ref{fig:sim}b, the electrical behavior for a network with a density of 1200 NWs/mm$^2$ (P regime) is depicted. Initially, it starts in a high-resistive state, and beyond a certain voltage threshold, it becomes activated, transitioning to a lower resistance state. As the voltage further increases, filament burning occurs, leading to higher resistance. Upon decreasing the voltage, filament formation restarts. If the current through the filament is low enough so the dissipated power is lower than the critical value P$_\mathrm{cw}$ the filament remains and the junction switches to the closed state. (Refer to Video 1). This simulation illustrates the activation process observed in percolated samples. 

In Figure \ref{fig:sim}c, the activation of filamentary junctions can be simulated through the application of a \SI{0.5}{\volt} voltage, resulting in a decrease in resistance. Subsequently, upon removing the voltage, these junctions begin to degrade, leading to an increase in electrical resistance. The resistance change upon relaxation can be compared to Figure \ref{fig:F2}b.

In Figure \ref{fig:sim}d, the electrical behavior for a network with a density of 2500 NWs/mm$^2$ (OP regime) is depicted. Initially, it starts in a state of very low linear resistance. While junction activation may occur, it does not imply a significant change in the network's resistance. When a critical power is reached, wire breakdown commences, leading to the disruption of percolative paths in the network. This reproduces the experimentally observed electro-fusing process. Upon conducting a second sweep (Figure \ref{fig:sim}e), filaments already formed in the previous cycle break upon reaching a certain power threshold. Filaments recover while reducing the applied voltage (Refer to Videos 2 and 3, Supporting Information). 
These results successfully reproduce those observed in over-percolated samples.

Correspondence with the experimental electrical response obtained using the two-junction model is notorious. On the one hand, by modifying the NWs' areal density, this phenomenological model successfully describes the percolation curve. On the other hand, besides the pristine resistance state, the electrical behavior of P and OP samples is also emulated. Additionally, the coexistence of two types of junctions serves as an explanation for the reversible switching identified around V$_{\mathrm{S}}$ in the experimental P case. In the OP case, the electro-fusing operation was also successfully simulated, and the different electrical responses found afterward agreed with those experimentally observed. 
\section{Conclusion}

\vspace{0.5cm}

In this work, the electrical response of AgNWNs has been characterized as a function of the NWs' areal density identifying two distinctive regimes referred to as percolated (P) and over-percolated (OP). While the former was extensively analyzed due to their promising features to implement neuromorphic-related properties, the latter had been exploited almost exclusively for its highly conductive linear I-V response. 
In this paper, we explore in detail the electrical behavior of the OP samples with switching properties to look for similarities and differences compared to their lower areal density coverage counterparts. We show that the first irreversible resistance change (termed electro-fusing) allows reducing the number of available low resistance paths, bringing the system closer to the percolation limit in an externally controlled way. 
This leads us to propose a unifying model to explain their salient features, i.e. the pristine resistance state and the ulterior switching behavior.

Based on the nature of the junctions formed at the crosspoint between NWs, a phenomenological two-type model is proposed. In this scenario, the NWNs are comprised of a robust permanent metallic-like junction and a labile switchable filamentary type. The relative population of the two-junction types determines the overall electrical behavior of AgNWNs samples as a function of the NWs' areal density. In particular, setting a high percentage of metallic-like junctions ($\sim 96 \%$) over the total amount (metallic + filamentary) successfully replicates the experimental percolation curve. With the set of parameters used to get the best agreement between simulated and experimental percolation data, the model effectively reproduces the posterior switching properties of the two highlighted types of samples (P and OP). 

In summary, by the combination of multiple AgNWNs fabrication strategies (including samples with different AgNWs geometries, and different NWs areal densities), and their detailed electrical characterization (considering the statistical analysis of the most salient features), we have collected multiple responses corresponding to different percolation regimes. Remarkably, considering very few assumptions (the main one being the nature of the two-junction types, which is strongly rooted in the origin and mobility of the Ag species within the NWNs), the proposed model serves to explain the overall behavior of P and OP samples both in their pristine state (percolation curve) and their ulterior switching phenomenology.     

Having demonstrated that the phenomenological model proposed in this work is plausible, from the microscopic point of view, and successful, in describing multiple samples in the different percolation regimes, it comprises a concrete step forward toward the deep understanding of self-assembled AgNWNs currently receiving an increasing interest for a broad spectrum of applications.

\section{Experimental Section}
Unless stated otherwise, micro-fabricated Ag pads were sputtered after deposition of the NWs, using a stencil mask (distance between electrodes is \SI{1}{\milli\meter}). The spread of the NWNs is thus restricted to an area of (1 mm)$^2$. External wires were attached to the electrodes using silver paste. Electrical measurements were performed (4-wire method by applying V with GW Instek GPP-2323 and measuring I and V with GW Instek GPD-8051 and GW Instek GPP-8062). The ambient relative humidity was around 33 $\%$ \cite{diaz_schneider_resistive_2022}. 
\medskip

\textbf{Supporting Information} \par 
Supporting material contains additional information to emphasize the broad validity of the resistive switching processes of percolated and over-percolated samples. That document: 1. expands the phenomenology of percolative switching samples (emphasizing the temporal dependence of the resistance), 2. includes a statistical analysis of the switching for multiple OP samples repetitively measured as well as measurements performed on the temporal domain, and 3. presents evidence that the reported switching is observable in AgNWNs regardless of the specific geometry of the constituent NWs.
Supporting Information is available from the Wiley Online Library or from the author.

\medskip
\textbf{Acknowledgements} \par 
The authors thank Dr. Carlos Acha (CONICET, UBA), Dr. Leticia Granja (INN, CONICET-CNEA), Dr. Ruben Weht (CNEA, UNSAM), Dr. Gustavo Segovia (INN, CONICET-CNEA), and Dr. Oscar Filevich (CONICET, UNSAM) for their insightful comments. We are also grateful to Dr. José Lipovetzky (CAB-CNEA) for technical assistance and Prof. Enrique Miranda for the valuable discussions on this work. We thank Dr. Carlos Bertoli (CAB-CNEA) for the SEM images. This work was supported by ANPCyT-FONCyT PICT 2021-0876 and PICT Start-Up 2019-00017. EDM acknowledges financial support from CONICET PIBAA 2022-2023 28720210100473CO. CPQ acknowledges financial support from CONICET PIBAA 2022-2023 28720210100975CO and PICT INVI 2022 01097. JIDS acknowledges a fellowship from CONICET.

\medskip

\bibliographystyle{MSP}
\bibliography{Bibliography.bib}

\newpage
\begin{figure}
\textbf{Table of Contents}\\
\medskip
 \includegraphics[width = 0.4 \textwidth]{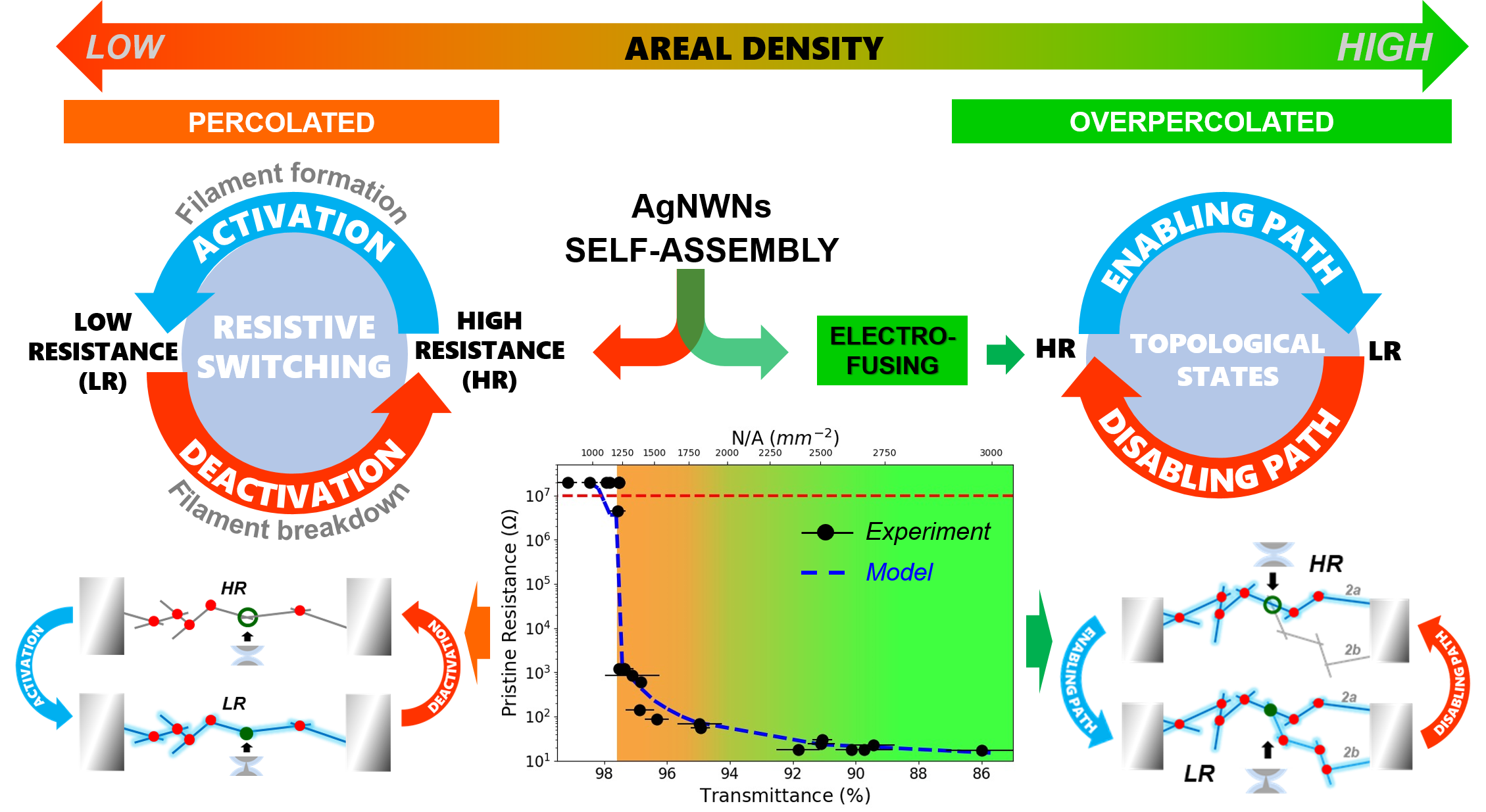}
  \medskip
  \caption*{A physical model with two types of junctions in silver nanowires is proposed to understand resistive switching in nanowire networks of any concentration.}
\end{figure}

\end{document}